# Calibration of Polarimetric Radar Data using the Sylvester Equation in a Pauli Basis

S. R. Cloude, AEL Consultants, Cupar, Fife, Scotland, UK, e-mail : aelc@mac.com

*Abstract*—In this paper we develop a new approach to the calibration of polarimetric radar data based on two key ideas. The first is the use of in-scene trihedral corner reflectors not only for radiometric and geometric calibration but also to completely remove any receiver distortion components. Secondly, we then show that the remaining transmitter distortion acts as a similarity transformation of the true scattering matrix. This leads us to employ a change of base to the Pauli matrix components. We show that in this basis calibration and the effects of Faraday rotation become much simplified and for example by using reciprocity alone we can then solve for copolar channel imbalance. Finally by using an uncalibrated symmetric point target of opportunity we can estimate cross-talks and hence fully solve the calibration problem without the need for using clutter averaging or symmetry assumptions in the covariance matrix as used in many other algorithms.

*Index Terms*—Polarimetric Calibration, Radar Polarimetry

## I. Introduction

The calibration problem for polarimetric radar data can be expressed succinctly as a matrix product model as shown in equation 1. Here [R] and [T] are 2x2 complex distortion matrices for receiver and transmitter respectively, [S] is the true 2x2 complex scattering matrix of the pixel and [N] is a noise matrix.

$$[S]_{obs} = [R][S][T] + [N]$$
$$= \begin{bmatrix} r_{11} & r_{12} \\ r_{21} & r_{22} \end{bmatrix} \begin{bmatrix} hh & hv \\ vh & vv \end{bmatrix} \begin{bmatrix} t_{11} & t_{12} \\ t_{21} & t_{22} \end{bmatrix} + [N] \quad - 1)$$

Many authors have worked with this model to derive practical calibration routines [1,2,3,4]. The most popular of these are based on use of a deployed trihedral corner reflector plus clutter averaging over a region where it is assumed that like and cross linear polarizations are uncorrelated. There are however two problems with these techniques:

1) The assumptions of reflection symmetry required for zero correlation are not guaranteed in a scene. Although some authors [4] have attempted to overcome this, one remaining issue is that the enforcement of this symmetry condition leads essentially to a zero coherence approximation. The estimation of low coherence in speckle environments (as found in coherent radar imaging) is notoriously difficult, with bias issues and a very large number of samples required for good estimation. For these reasons it would be better to find alternative strategies where possible.

2) The deployed trihedrals are used traditionally for three purposes, radiometric and geometric calibration as point scatterers of known location and backscatter cross section. They are also used for quadpol data to estimate a single complex number, the copolar channel imbalance as part of the calibration scheme [3]. However, as shown in [5,6] and to be further developed here, such reflectors have more degrees of freedom available to help in the calibration process.

In this paper we develop a new approach to calibration that avoids the need for speckle averaging of covariance data and also makes full use of information from deployed trihedral reflectors. In section II we describe the basic algorithm before showing in section III how use of the Pauli matrix basis yields simplified equations for calibration. In section IV we show details of the calibration scheme before considering the special but important case of Faraday rotation distortions in section V. Finally in section VI we show an example of application of this new technique to simulated POLSAR data.

## II. Trihedrals and the Sylvester Equation

We start by noting that trihedrals have the general property of having the identity matrix as an ideal reference matrix. In practice therefore the measured [S] matrix for each trihedral has the form shown in equation 2, i.e. a product of receiver and transmitter distortion matrices.

$$[S] = \begin{bmatrix} 1 & 0 \\ 0 & 1 \end{bmatrix} \Rightarrow [S_{CR}]_{obs} = [R][T] \quad - 2)$$

This may not seem too helpful until we make use of the inverse of the measured matrix to remove completely the receiver distortion matrix, as shown formally in equation 3 and first proposed in [5] as a free space measurement prior to full system calibration.

$$[S]_{obs} = [R][S][T] \Rightarrow [S_{CR}]_{obs}^{-1}[S]_{obs}$$
$$= [T]^{-1}[R]^{-1}[R][S][T] = [T]^{-1}[S][T] \quad - 3)$$

Note that we can also choose to remove the transmitter by post multiplication and that in the presence of noise [N] we can measure the individual matrices for a deployed set of trihedrals and then form the estimate of [$S_{CR}$] from the dominant eigenvector of the covariance matrix of corresponding scattering vectors.

### III. PAULI MATRIX EXPANSION OF THE SYLVESTER EQUATION

Equation 3 is a standard form of the Sylvester equation from control theory [5,6]. In essence we must use observed matrices to solve for an unknown matrix [T], which we know only distorts a matrix [S] by a similarity transformation. This knowledge does however give us a clue how to simplify the problem. The Pauli matrix expansion is widely used in polarimetry [7], largely because it leads to invariants under similarity transformations (roll or rotation invariance for example). With this in mind we first vectorize equation 3 as shown in 4 (where we drop brackets around matrices for notational convenience).

$$S_{obs} = T^{-1}ST \Rightarrow \underline{s}_{obs} = T^{-1} \otimes T^T \underline{s} \quad - 4)$$

with $\underline{s} = [hh\ hv\ vh\ vv]^T$. We now convert this into the Pauli basis by forming sums and differences of components as shown in 5 (note that from now on all $\underline{k}$ vectors are obtained after multiplying the observed S matrix data by the inverse of the corner reflector matrix).

$$\underline{k} = \frac{1}{\sqrt{2}} \begin{bmatrix} 1 & 0 & 0 & 1 \\ 1 & 0 & 0 & -1 \\ 0 & 1 & 1 & 0 \\ 0 & 1 & -1 & 0 \end{bmatrix} \begin{bmatrix} hh \\ hv \\ vh \\ vv \end{bmatrix} = [U_P]\underline{s} \quad - 5)$$

$$\Rightarrow \underline{k}_{obs} = [A_P]\underline{k} = U_P T^{-1} \otimes T^T U_P^{-1} \underline{k}$$

Working through the algebra using equations 1 and 3 we obtain the following key result, that the remaining distortion is given by a 4x4 sparse matrix [A] as shown in equation 6. Note that here we have used two key simplifying ideas, first the weak coupling assumption so that in magnitude $t_{11}, t_{22} \gg t_{12}, t_{21}$ and reciprocity in the backscatter direction to simplify [$A_P$] from equation 5 to generate [A], the fourth column of which is set to $[0\ 0\ 0\ 1]^T$ (we can do this because the distortion by a given T is independent of this 4th column vector, since the 4th element of $\underline{k}$ (the true underlying scattering vector for a pixel) is always zero). This makes the calibration matrix easier to invert as will be shown in equation 9. Note that inversion of the full matrix [$A_P$] is required for general bistatic scattering problems, as found in optical systems for example [5,6]. Here we can make good use of the reciprocity theorem and weak coupling assumption to simplify the problem of calibration for monostatic radar systems and use the sparse matrix [A] in equation 6. The parameters in [A] are defined as shown in 7 and 8. Equation 7 shows two copolar ratio terms, m and c, while equation 8 shows three crosspolar ratios.

$$\underline{k}_{obs} = [A]\underline{k} \qquad [A] = m \begin{bmatrix} 1 & 0 & 0 & 0 \\ 0 & 1 & x_1 & 0 \\ 0 & x_2 & C_1 & 0 \\ 0 & x_3 & C_2 & 1 \end{bmatrix} \quad - 6)$$

$$m = \frac{t_{11}t_{22}}{t_{11}t_{22} - t_{12}t_{21}} \approx 1 \quad c = \frac{t_{22}}{t_{11}} \Rightarrow \begin{cases} C_1 = \frac{1}{2}(c + \frac{1}{c}) \\ C_2 = \frac{1}{2}(c - \frac{1}{c}) \end{cases} \quad - 7)$$

$$x_1 = \frac{t_{21}}{t_{11}} - \frac{t_{12}}{t_{22}} = \frac{x_3(c^2 - 1) - x_2(c^2 + 1)}{2c}$$
$$x_2 = \frac{t_{12}}{t_{11}} - \frac{t_{21}}{t_{22}} \quad x_3 = \frac{t_{12}}{t_{11}} + \frac{t_{21}}{t_{22}} \quad - 8)$$

### IV. CALIBRATION: SOLVING THE SYLVESTER EQUATION

We see from equation 6 that the HH+VV channel is already calibrated by the corner reflector (the factor m is close to unity for small coupling). To fully correct the data, calibration now reduces to matrix inversion, which can be easily evaluated as shown in equation 9.

$$\underline{k} = [A]^{-1}\underline{k}_{obs}$$

$$[A]^{-1} = \frac{1}{m(C_1 - x_1 x_2)} \begin{bmatrix} C_1 - x_1 x_2 & 0 & 0 & 0 \\ 0 & C_1 & -x_1 & 0 \\ 0 & -x_2 & 1 & 0 \\ 0 & a_{42} & a_{43} & C_1 - x_1 x_2 \end{bmatrix} \quad - 9)$$

$$a_{42} = x_2 C_2 - C_1 x_3 \qquad a_{43} = x_1 x_3 - C_2$$

To fill this matrix we can first use reciprocity, which states that [S] is complex symmetric and so the fourth element of $\underline{k}$ must be zero. If $\underline{k}_{obs} = [z_0\ z_1\ z_2\ z_3]^T$ then this leads to the following equation

$$\frac{a_{42}}{m(C_1 - x_1 x_2)} z_1 + \frac{a_{43}}{m(C_1 - x_1 x_2)} z_2 + \frac{1}{m} z_3 = 0 \quad - 10)$$

By selecting N > 1 points we can then use this equation to estimate two calibration parameters, $a_{42}$ and $a_{43}$, as shown in equation 11 (if N>2 then a least squares solution is implicit in 11). From these we can then obtain two primary parameters, the transmitter complex copolar imbalance from $a_{43}$ and an 'average' cross-talk term from $a_{42}$ as shown in 12. Note that we have a sign ambiguity in the copolar ratio. Equation 12 gives partial information about cross-talks (and as we shall see an indication of any Faraday rotation in trans-ionospheric propagation) but in order to determine these explicitly requires access to point targets with zero cross-polarization and unequal copolar scattering coefficients as shown in equation 13.

$$\frac{1}{C_1-x_1x_2}\begin{bmatrix} z_1^1 & z_2^1 \\ z_1^2 & z_2^2 \\ \vdots & \vdots \\ z_1^N & z_2^N \end{bmatrix}\begin{bmatrix} a_{42} \\ a_{43} \end{bmatrix} = \begin{bmatrix} -z_3^1 \\ -z_3^2 \\ \vdots \\ -z_3^N \end{bmatrix} \quad -11)$$

$$= \frac{1}{C_1-x_1x_2}[Z]\begin{bmatrix} a_{42} \\ a_{43} \end{bmatrix} = \underline{b} \Rightarrow \frac{1}{C_1-x_1x_2}\begin{bmatrix} a_{42} \\ a_{43} \end{bmatrix} = Z^+\underline{b}$$

$$\frac{a_{42}}{C_1-x_1x_2} = \frac{x_2C_2-C_1x_3}{C_1-x_1x_2} \approx \frac{C_2}{C_1}x_2 - x_3$$

$$\frac{a_{43}}{C_1-x_1x_2} = \frac{x_1x_3-C_2}{C_1-x_1x_2} \approx -\frac{C_2}{C_1} = \varepsilon = \frac{1-c^2}{1+c^2} \quad -12)$$

$$\Rightarrow c = \pm\sqrt{\frac{1-\varepsilon}{1+\varepsilon}} = \pm\frac{t_{22}}{t_{11}}$$

$$[S] = \begin{bmatrix} a & 0 \\ 0 & b \end{bmatrix} \quad a \neq b \in C \quad -13)$$

Here a,b can be unknown and so we do not need to use a radiometrically calibrated point target. The observed scattering vector for this target is then determined as shown in 14.

$$\begin{bmatrix} z_0 \\ z_1 \\ z_2 \\ z_3 \end{bmatrix} = \begin{bmatrix} a+b \\ a-b \\ x_2(a-b) \\ x_3(a-b) \end{bmatrix} \Rightarrow \begin{cases} x_2 = \frac{z_2}{z_1} \\ x_3 = \frac{z_3}{z_1} \end{cases} \quad -14)$$

Note that the best target to use would be a metallic dihedral (a=-b) and these could be deployed for calibration purposes but notice that their RCS and detailed scattering matrix need not be known and so this relieves the engineering requirements on such reflectors. Also any suitable non-trihedral of opportunity could also be used if available in the scene (urban areas for example), although details of the best algorithm to merge available point targets remain to be developed in the future.

From 14 we see that we can then directly estimate $x_2$ and $x_3$ and hence from equation 8, $x_1$. This allows us to fully populate the calibration matrix as shown in equation 15 (and if we wish, to reconstruct the R and T distortion matrices).

$$\hat{A}^{-1} \approx \begin{bmatrix} 1 & 0 & 0 & 0 \\ 0 & 1 & -\frac{x_1}{C_1} & 0 \\ 0 & \mp\frac{x_2}{C_1} & \pm\frac{1}{C_1} & 0 \\ 0 & \frac{a_{42}}{C_1} & \frac{a_{43}}{C_1} & 1 \end{bmatrix} \quad C_1 = \pm\frac{1+c^2}{2c} \quad -15)$$

This matrix can then be used to correct all distortions of the data and provide full polarimetric calibration. Note that the sign ambiguity in 12 translates into a sign ambiguity in the third row of this matrix and so the only consequence is a sign change in the final calibrated cross-polar channel.

V. FARADAY ROTATION AND THE SYLVESTER EQUATION

One interesting special case of the general distortion model is for space-borne low frequency sensors when the transmitter has copolar imbalance but zero system crosstalk, which is then combined with Faraday rotation due to trans-ionospheric propagation. This rotation, due to its non-reciprocal nature, then appears like an effective cross-talk term in classical approaches. Such a model for example was found appropriate for the Japanese L-band radar satellite ALOS-1. Mathematically this model has the form shown in 16, where $\theta_f$ is the one-way Faraday rotation

$$[T] = [R(\theta_f)][T_D] = \begin{bmatrix} \cos\theta_f & -\sin\theta_f \\ \sin\theta_f & \cos\theta_f \end{bmatrix}\begin{bmatrix} t_{11}^d & 0 \\ 0 & t_{22}^d \end{bmatrix} \quad -16)$$

One problem we face here is that $\theta_f$ is scene dependent and can be large (especially for proposed P-band satellites such as the ESA-BIOMASS mission) and so the small coupling assumptions made in classical calibration analyses can become invalid. Vectorizing equation 16 leads to the matrix form shown in equation 17. Here we see a special form of the distortion matrix of equation 6.

$$\underline{k}_{obs} = A\underline{k} = \begin{bmatrix} 1 & 0 & 0 & 0 \\ 0 & 1 & 0 & 0 \\ 0 & 0 & C_1 & 0 \\ 0 & 0 & C_2 & 1 \end{bmatrix}\begin{bmatrix} 1 & 0 & 0 & 0 \\ 0 & \cos 2\theta_f & -\sin 2\theta_f & 0 \\ 0 & \sin 2\theta_f & \cos 2\theta_f & 0 \\ 0 & 0 & 0 & 1 \end{bmatrix}$$

$$\underline{k} = \cos 2\theta_f \begin{bmatrix} \sec 2\theta_f & 0 & 0 & 0 \\ 0 & 1 & -\tan 2\theta_f & 0 \\ 0 & C_1\tan 2\theta_f & C_1 & 0 \\ 0 & C_2\tan 2\theta_f & C_2 & \sec 2\theta_f \end{bmatrix}\underline{k}$$

$$-17)$$

Equation 18 shows this relationship more explicitly and demonstrates a key result. Even though the distortion matrix looks to contain effective cross-talk (given by $\tan 2\theta_f$), the inverse matrix solution shows a zero $a_{42}$ element (using equation 12) i.e. it correctly identifies zero system cross-talk. The $a_{43}$ element again correctly shows the copolar imbalance.

$$\begin{bmatrix} m & 0 & 0 & 0 \\ 0 & 1 & x_1' & 0 \\ 0 & x_2' & C_1' & 0 \\ 0 & x_3' & C_2' & m \end{bmatrix} \Rightarrow \begin{cases} a_{43} = -\frac{C_2'}{C_1'} = \varepsilon \\ \frac{a_{42}}{C_1'-x_1'x_2'} = 0 \end{cases} \quad -18)$$

The physical reason for this is that Faraday rotation is converted by the trihedral inverse matrix multiplication into an ordinary rotation. In this way Faraday rotation is no longer confused with system cross-talk. Note that this is also true for an arbitrary distortion matrix T, as shown in equation 19. Faraday rotation can therefore be separated from the matrix T in the calibration process, irrespective of the magnitude of rotation. However it does cause an 'ordinary' rotation of every pixel (which complicates the simple cross-talk estimation algorithm in equation 14).

$$T_f = R(\theta_f)T \rightarrow S_{obs} = T_f^{-1}ST_f$$
$$= T^{-1}R(-\theta_f)SR(\theta_f)T = T^{-1}S(\theta_f)T \quad - 19)$$

The resulting rotation of [S] can then be mitigated either by using roll invariant processing (entropy/alpha for example [7]) or by standard algorithms for rotation compensation in radar polarimetry.

## VI. AN EXAMPLE

We now illustrate the above algorithms by using a simulated L-band POLSAR image obtained from PolsarproSim, the fully polarimetric simulator embedded in the ESA-Polsarpro freeware software package (http://earth.eo.esa.int/polsarpro/ ). Range and azimuth resolutions were set to 1.4/0.7m, typical of current high-resolution SAR systems. The pixel size is 0.5m. We simulate a small patch of 10m high forest surrounded by rough surface scattering, where we locate a 1m trihedral reflector. We then add the following receiver [R] and transmitter [T] distortion matrices to the data using equation 1.

$$[R] = \begin{bmatrix} 0.7989 & -0.0419 \\ -0.0556 - 0.0149i & -1.0611 - 0.2843i \end{bmatrix}$$
$$[T] = \begin{bmatrix} 0.9986 & -0.0589 - 0.034i \\ 0.0523 & 1.1243 + 0.6491i \end{bmatrix} \quad - 20)$$

Figure 1 shows the resulting uncalibrated scattering matrix data (processed with a 3x3 boxcar speckle filter). Here we see that the corner reflector has 'leaked' into all four channels and that the forest canopy appears not only in HV+VH but also in HV-VH and hence violates reciprocity.

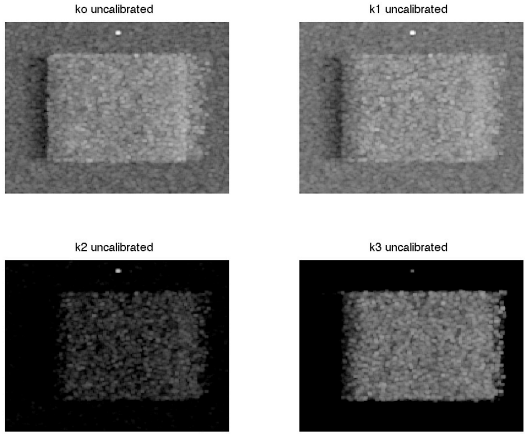

Figure 1: Simulated L-band SAR scene of forest and corner reflector showing four Pauli matrix channels of uncalibrated data. $k_0$ = |HH+VV|, $k_1$ = |HH-VV|, $k_2$ = |HV+VH| and $k_3$ = |HV-VH|

By isolating the SLC data around the corner reflector we obtain an estimate of the product scattering matrix and its inverse as shown in equation 21. Figure 2 shows the result of applying this inverse matrix to each pixel of the SLC image. Here we see that the corner reflector has been isolated in the HH+VV channel as it should be and indeed the $k_0$ image is now fully calibrated without further processing.

$$[S_{CR}] = \begin{bmatrix} 1 & -0.1183 - 0.0684i \\ -0.1396 - 0.0375i & -1.2633 - 1.265i \end{bmatrix}$$
$$\quad - 21)$$
$$[S_{CR}]^{-1} = \begin{bmatrix} 1 & -0.0738 + 0.0198i \\ -0.07 + 0.0404i & -0.3953 + 0.3958i \end{bmatrix}$$

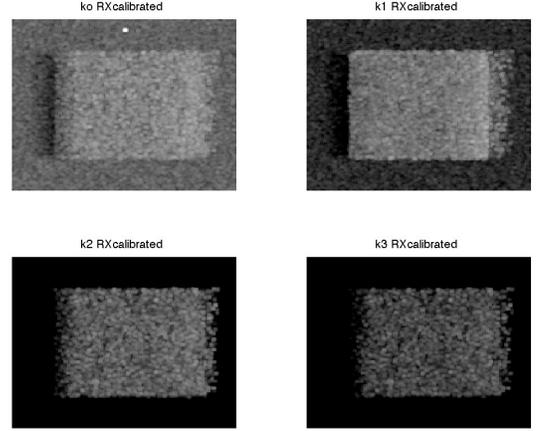

Figure 2: Receiver calibrated Pauli channels using the inverse matrix of equation 21

However we see that there is still some HV-VH signal and hence reciprocity is violated.

$$A^{-1} = \begin{bmatrix} 1 & 0 & 0 & 0 \\ 0 & 1 & 0 & 0 \\ 0 & 0 & 1.09 - 0.1619i & 0 \\ 0 & 0 & -0.3349 - 0.5281i & 1 \end{bmatrix} \quad - 22)$$

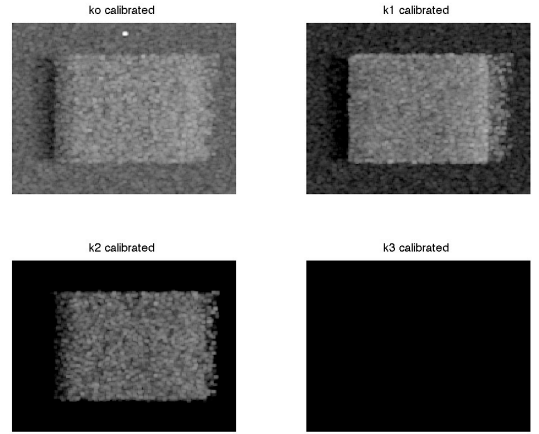

Figure 3: Reciprocity calibrated Pauli matrix channels using the matrix of equation 22

By enforcing this using equation 10 for the forest pixels we obtain estimates of the copolar channel imbalance and cross-

5talk as shown in the inverse matrix of equation 22 and applied in figure 3. We note that this matrix contains zero apparent system cross-talk. This is correct, as the original distortion matrices (equation 20) were obtained from the model of equation 16 with 3 degrees of Faraday rotation. Hence we can confirm that the data in figure 3 is now fully calibrated using only the trihedral and reciprocity

## VII. Conclusion

In this paper we have developed a new approach to the calibration of polarimetric radar data. We first made full use of information from deployed trihedrals to remove receiver distortion before showing how reciprocity and uncalibrated dihedrals can then be used to perform calibration in the Pauli matrix basis. We further showed how the procedure transforms Faraday rotation into an 'ordinary' rotation of the data and hence separates it from apparent system cross-talk. We believe that the procedure greatly simplifies the calibration of quadpol data and have demonstrated the technique using a simulated SAR scene from Polsarpro.